\documentclass[twocolumn]{aastex631}

\usepackage{xspace}
\usepackage{enumerate}
\usepackage{lineno}

\newcommand{\yr}    	{\ifmmode \mathrm{yr} \else yr\fi}
\newcommand{\mpc}   	{\ifmmode \,\mathrm{Mpc}^{-3} \else \,Mpc$^{-3}$\fi}
\newcommand{\Msun}	    {\ifmmode \,\mathrm M_{\odot} \else $\,\mathrm M_{\odot}$\fi\xspace}
\newcommand{\Zsun}	    {\ifmmode \,\mathrm Z_{\odot} \else $\,\mathrm Z_{\odot}$\fi\xspace}
\newcommand{\Mhalo} 	{\ifmmode M_{\mathrm{halo}} \else $M_{\mathrm{halo}}$\fi\xspace}
\newcommand{\Rvir}  	{\ifmmode R_{200} \else $R_{200}$\fi\xspace}
\newcommand{\Mstar}	    {\ifmmode {M}_{\star} \else ${M}_{\star}$\fi\xspace}
\newcommand{\Mvir}	    {\ifmmode M_{\mathrm{halo}} \else $M_{\rm halo}$ \fi\xspace}
\newcommand{\Htwo}  	{\ifmmode {\rm H}_{2} \else ${\rm H}_{2}$ \fi}
\newcommand{\nH}    	{\ifmmode {n}_{\rm H} \else ${n}_{\rm H}$ \fi}
\newcommand{\cc}	    {\ifmmode {\rm cm}^{-3} \else ${\rm cm}^{-3}$ \fi}
\newcommand{\mperyr}	{\ifmmode \Msun{\rm yr}^{-1} \else $\Msun{\rm yr}^{-1}$ \fi}
\newcommand{\arepo}	    {{\small AREPO}\xspace}

\newcommand{\civ}	    {\ifmmode {\rm C}_{\rm IV} \else CIV \fi}
\newcommand{\mgii}	    {\ifmmode {\rm Mg}_{\rm II} \else MgII \fi}
\newcommand{\oi}	    {\ifmmode {\rm O}_{\rm I} \else OI \fi}
\newcommand{\perccm}	{\ifmmode {\rm cm}^{-2} \else ${\rm cm}^{-2}$ \fi}

\shorttitle{Cosmological GC formation}
\shortauthors{Gutcke et al.}
\begin{document}

\title{Low--mass globular clusters from stripped dark matter halos}

\email{gutcke@hawaii.edu}
\author[0000-0001-6179-7701]{Thales A. Gutcke}\thanks{NASA Hubble Fellow}
\affiliation{Institute for Astronomy, University of Hawaii, 2680 Woodlawn Drive, Honolulu, HI 96822, USA}

\begin{abstract}

The origin and formation of globular clusters has remained a mystery. We present a formation scenario for ancient globular cluster-like objects that form in ultra-high resolution simulations (smallest cell size $<0.1\,\mathrm{pc}$, mass resolution $M_\mathrm{cell}=4\,\Msun$). The simulations are cosmological zoom-in simulations of dwarf galaxies within the stellar mass range $10^{6-7}\,\Msun$ that match Local Group dwarf properties well. Our investigation reveals globular clusters hosting ancient stellar populations, characterized by a lack of dark matter in the present epoch. The clusters exhibit short, episodic star formation histories, occasionally marked by the presence of multiple stellar generations. The metallicity distributions show a widening, encompassing stars in the range of $10^{-4} < Z_\star/\Zsun < 1$. The presence of these objects is attributable to star formation occurring within low-mass dark matter halos ($\Mhalo\approx10^6\,\Msun$) during the early stages of the Universe, preceding Reionization ($z\gtrsim7$). As these clusters are accreted into dwarf galaxies, dark matter is preferentially subjected to tidal stripping, with an average accretion redshift of $\bar{z} \approx 5$.

\end{abstract}
\keywords{}

\section{Introduction} \label{sec:intro}

The genesis of globular clusters (GCs) remains a subject of intense scrutiny within the astrophysical community. The question of whether GCs formed within dark matter halos or emerged from gravitationally bound clouds in the early universe remains a topic of ongoing debate \cite[e.g.][]{Peebles1968, Searle1978, Kravtsov2005, Penarrubia2017}. GCs manifest across diverse galactic environments and exhibit a linear relationship between total GC mass and inferred halo mass, a relationship that holds remarkably consistent across a wide range of scales \citep{Spitler2009, Forbes2018}. This contrasts with the stellar mass - halo mass relation, characterized by multiple power laws with varying exponents \cite[e.g.][]{Moster2013}. This suggests a formation mechanism more closely tied to the buildup of dark matter halos than to the subsequent processes of star formation within them. Furthermore, it appears distinct from mechanisms driving baryonic concentration within halos and subsequent star formation in galactic components like disks, bulges, and spheroids, as well as those quenching global levels of star formation, such as stellar and AGN feedback. If indeed disconnected from these processes, GC formation challenges current state-of-the-art models \cite[e.g.][]{Kravtsov2005, Li2017, Pfeffer2018}.

\cite{ElBadry2019} posit that the GC mass--halo mass relationship for $M_\mathrm{halo} > 10^{11.5}\Msun$ may be a result of hierarchical assembly, an outcome of the central limit theorem. However, \cite{Forbes2018} extend the observed relationship down to $M_\mathrm{halo} \approx 10^{9}\Msun$. They note that the scatter of the relation increases toward lower masses, partially attributed to larger uncertainties in halo mass estimates for these galaxies. Given that dwarf galaxies at this lower mass range likely experienced fewer mergers, the hierarchical assembly interpretation becomes less plausible. As we shall demonstrate, a scenario where early luminous mini-halos are stripped of their dark matter more readily accounts for the GC mass-halo mass relationship.

An open question revolves around the minimum halo mass capable of hosting a GC. Presently, Eridanus II stands as the galaxy with the lowest known mass hosting a single GC. Remarkably, it aligns well with the GC mass--halo mass relationship, as its lone GC possesses a mass of $M_\mathrm{GC}\approx4\times10^3~\Msun$, an extremely low value compared to the general population of GCs. The estimated halo mass of Eridanus II is around $\Mhalo\approx 10^9~\Msun$.  Since the GC in Eridanus II sits at the lower end of the GC mass function and fits the GC mass --halo mass relation, this may indicate that smaller GCs may better withstand the conditions in lower mass galaxies. Additionally, \cite{Hayashi2023} reveal that the best-fitting dark matter halo profile is not cored, but rather peaked or ``cuspy.'' Cuspy profiles are conventionally thought to lead to more efficient tidal disruption of clusters, making the presence of a GC even more puzzling. However, this is modulated by the increasing fraction of dwarf galaxies devoid of GCs as halo mass decreases. 

While GCs can be found in many galaxies, the Milky Way's GC system remains the best studied. The GC mass (or luminosity) function in the Milky Way adheres to a power law with a turnover at the low mass end around $4.4\times10^4\,\Msun$ \citep{DiCriscienzo2006}. This observation can be understood by assuming a non-truncated functional form at the time of formation and subsequently invoking destruction mechanisms that reduce the numbers of GCs at the lower mass end. \cite{Baumgardt2003} conducted simulations of the dissolution of GCs in a tidal field, incorporating multi-mass stars following the initial mass function (IMF). Initially, mass loss of the GCs is primarily due to stellar evolution, resulting in a loss of approximately 30\% of their mass. Subsequently, tidal disruption within the Milky Way's potential and two-body evaporation \citep[following][]{Spitzer1987} assume prominence as the principal mass loss processes.

Indeed, GCs in dwarf galaxy environments are not so dissimilar. \cite{Georgiev2009} present a sample of GCs in dwarf galaxies based on HST/ACS observations. Employing the fitting functions from \cite{Baumgardt2001}, they argue that the absence of low mass GCs in their sample can also be well explained by considering evaporation and tidal disruption within the halos of these dwarfs. In a more recent survey, \cite{Carlsten2022a} investigate the properties of GCs in dwarf satellites of Milky Way-sized galaxies as part of the Exploration of Local VolumE Satellites (ELVES) Survey \citep{Carlsten2022b}. The authors constrain the occupation fraction of GCs in dwarf galaxies, highlighting the significant role of environment. Lower density environments tend to host dwarf satellites with a lower GC occupation compared to denser environments like the Virgo cluster. So, while denser environments may dissolve their GCs more readily, GCs are also much most likely to be accreted in such environments. 

Another interesting aspect of GCs is that they exhibit a pronounced dichotomy in color and metallicity, leading to their classification into distinct categories: old and young clusters. The former, typically associated with faint galaxies, possess low specific frequencies, while the latter, prevalent in massive galaxies and mergers, display higher specific frequencies. This distinction suggests that old GCs may trace their origins to the nascent stages of proto-galactic collapse, whereas younger counterparts arise in more turbulent, extreme environments \cite[e.g.][]{Kissler-Patig1997}. However, the precision of age determinations for ancient GCs is subject to significant uncertainties, ranging from their formation epoch at $z \approx 3$ to the possibility of extending into the epoch of reionization at $z_\mathrm{form} > 6$. Recent observations of a gravitationally lensed galaxy utilizing the James Webb Space Telescope (JWST) have unveiled a compelling case study: the "Sparkler," a galaxy at redshift $z=1.38$, hosting a population of mature globular clusters \citep{Mowla2022}. Analysis of these clusters indicate a formation epoch predating $z=9$, aligning with predictions from our simulations.

In addition to their age diversity, GCs often exhibit multiple stellar populations. Without a formation scenario involving dark matter halos, it becomes challenging to account for the emergence of distinct populations within a single GC. This challenge arises from the fact that the first generation of stars will disperse its natal cloud. To reaccrete sufficient gas to facilitate subsequent generations of stars without the gravitational potential of a dark matter halo necessitates invoking improbable physics or dynamical interactions with molecular clouds within a galaxy \citep[see the review by][]{Bastian2018}.

In recent years, significant progress has been made in the field of globular and star cluster formation through the use of simulations \cite[e.g.][]{Pfeffer2018, Kruijssen2019, RamosAlmendares2020, Doppel2021}. Some studies have employed cosmological simulations combined with semi-analytical or sub-grid models to incorporate clusters into the simulations after the fact. However, since these simulations do not resolve clusters themselves, they utilize a technique called `particle tagging' to assign sub-grid properties to stellar or dark matter particles. These models, calibrated to observed relations, are somewhat limited in their ability to probe potential formation mechanisms. In the same vein, the recent study by \cite{Grudic2023} utilizing the FIRE-2 model on Milky Way-mass galaxies employed post-processing to synthesize cloud properties and predict cluster characteristics using the STARFORGE model \citep{Grudic2021}. However, this simulation did not yield a population of ancient globular clusters. The authors posit that insufficient star formation before redshift $z=3$ in the simulations may account for this absence. 

In contrast, high-resolution simulations, both non-cosmological and run to high redshifts, have successfully modelled the formation of resolved star clusters, albeit not globular clusters per se \citep{Lahen2020, Hislop2022,Sameie2023}. The authors suggest that these star clusters might serve as proto-globular clusters, given adequate time to evolve. Nevertheless, the link between high-redshift formation and low-redshift properties remains unverified, and these clusters have not yet been confirmed to host multiple stellar populations or adhere to the GC mass-halo mass relation. 

This study focuses on the formation of present-day GC-like objects that are associated with dwarf galaxies and that display ancient populations. In particular, we will show that stripped DM halos can account for the GC mass--halo mass relation and provide a simple avenue to form multiple stellar populations in GCs. 

The paper is organized as follows: in Sec.~\ref{sec:model} we summarize the most relevant aspects of the simulation model. Sec.~\ref{sec:results} shows our main results from the simulations, comparing to relevant observations where available. Sec.~\ref{sec:discussion} discusses the assumptions and caveats to our results. Finally, Sec.~\ref{sec:conclusion} summarizes our major findings. Throughout this study, we assume the $\Lambda$CDM cosmological parameters $\Omega_{\Lambda}=0.693$, $\Omega_0=0.307$, $\Omega_b=0.048$, and h=0.6777 \citep{PlanckCollaboration2014}. For metallicities, we assume the solar value of $Z_{\odot}=0.01337$, from \cite{Asplund2009}.

\section{Model \& Method} \label{sec:model}

In this study, we employ a sample of five cosmological simulations run up to redshift $z=0$, utilizing the LYRA galaxy formation model \citep{Gutcke2021, Gutcke2022a}. These simulations feature dwarf galaxies identical to those in \citealt{Gutcke2022b}, where it has been demonstrated that their stellar properties at $z=0$ closely align with those of Local Group dwarf galaxies. This includes characteristics such as stellar mass, size, kinematics, metallicity, and star formation history. Additionally, we introduce a supplementary simulation (HaloF) conducted using the same model. These simulations encompass virial masses ranging from $8\times10^8\Msun$ to $9\times10^9\Msun$ at $z=0$, with corresponding stellar masses falling within the range of $3\times10^6\Msun$ to $10^7\Msun$. Except for two cases, the galaxies persist in forming stars at low rates between $10^{-4}$ and $10^{-6}\mperyr$ at $z=0$.

The LYRA model is a comprehensive numerical framework that incorporates a resolved interstellar medium (ISM) with a cooling prescription operative down to temperatures as low as $10$K. It further encompasses individual, star-by-star star formation, individually resolved supernovae events, and a subgrid model accounting for Population III (PopIII) star enrichment during the high-redshift era. These model prescriptions are implemented within the cosmological, hydrodynamical moving-mesh code \arepo \citep{Springel2010, Pakmor2016, Weinberger2020}. For an extensive description of the model and code characteristics, we refer the reader to the cited papers. In the ensuing discussion, we will outline the salient aspects most pertinent to this study.

The simulations are initialized with zero metallicity, precluding any initial metal cooling. Adopting the ``$10^6$'' case as presented in \cite{Gutcke2022a} as our fiducial model, we incorporate the enrichment by PopIII stars in a sub-grid fashion. Running the halo finder \texttt{Subfind} \citep{Springel2001} in real-time to compute the virial radius, the metallicity of all gas within this radius undergoes augmentation from $0$ to $10^{-4} \Zsun$ once a halo surpasses the mass threshold of $M_\mathrm{PopIII}=10^6\Msun$. Subsequently, when the gas metallicity reaches $10^{-4} \Zsun$ or higher, it is permitted to cool to temperatures of $T\geq10~\mathrm{K}$ and condense to $n_\mathrm{H} \leq 10^4~\mathrm{cm}^{-3}$--this is where star formation takes place. As demonstrated in \cite{Gutcke2021}, stars commence formation in halos with $M\mathrm{halo}\geq10^6~\Msun$, where supernovae yield metal-enriched outflows. These outflows, in turn, disseminate metals to adjacent, less massive halos, increasing the cooling rate and, subsequently, triggering star formation. Consequently, stars may form in halos with masses as low as $M\mathrm{halo}\gtrsim10^3~\Msun$.

The zoom-in initial conditions are generated from the EAGLE simulation \citep{Schaye2015}, following the methodology outlined in \cite{Jenkins2013}. To create these conditions, the complete EAGLE box ($L=100,h^{-1}$~Mpc) is simulated at low resolution for each halo. Within this box, we define a high-resolution region encompassing the entire Lagrangian region of a single dwarf galaxy. Within this Lagrangian region, the gas mass resolution is set at $4\,\Msun$. To maintain consistent mass resolution as the simulation progresses, cells can be refined and de-refined. The dark matter (DM) mass resolution is approximately $80\,\Msun$. In our "high-resolution" run, employed for examining the size-mass relation (see Sec.~\ref{sec:size}), the gravitational softening length is $1$~pc for DM, gas, and stars. For the rest of the simulations, the softening lengths are $4$~pc for gas and stars, and $10$~pc for DM. The selection of galaxies adheres to an isolation criterion, ensuring they do not interact with larger galaxies throughout their lifespan.

\section{Results} \label{sec:results}

\begin{figure*}
    \centering
    \includegraphics[width=\textwidth]{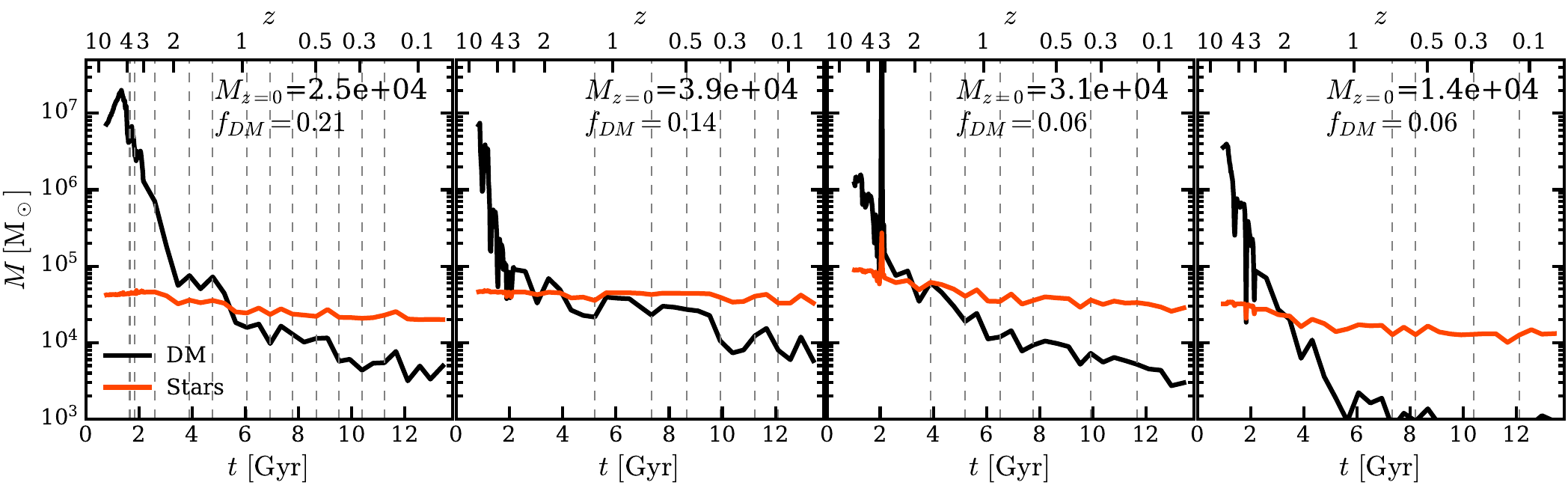}
    \caption{The mass evolution of four stripped halos. Black lines show the DM, red lines the stellar mass at a function of time. Grey vertical lines show pericenter passages during infall into the main halo. Most stripped halos have halo masses of $\Mhalo\gtrsim10^6\Msun$ at the time when their stars form and before they are stripped of their DM. Stellar mass dominates after around $z\approx2$. Most halos are stripped of their DM before they fall into our primary galactic halo.}
    \label{fig:gridmass}
  \end{figure*}

   \begin{figure}
    \centering
    \includegraphics[width=\columnwidth]{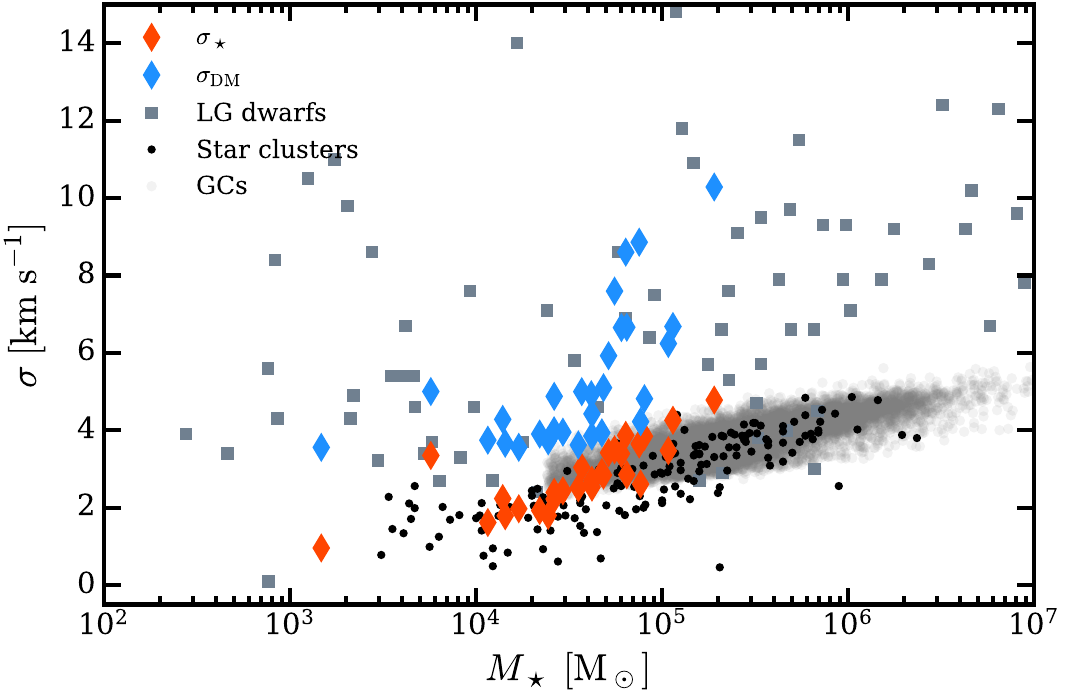}
    \caption{Velocity dispersion of halos at $z=5.5$, a period of time before the majority of DM stripping has occurred. Red diamonds show the stellar velocity dispersion and blue show the dispersion in the dark matter. For comparison, we also show the measurements for star clusters \citep{McLaughin2005}, globular clusters \citep{Harris2020}, and Local Group dwarf galaxies \citep{McConnachie2012}. The DM is preferentially stripped because it has a higher velocity dispersion relative to the stars.}
    \label{fig:sigmaz5}
  \end{figure}

In this paper, we consider the substructure found within the virial radius at $z=0$. The virial radii at this time are $18 -35~\mathrm{kpc}$. We detect substructure using the code \texttt{Subfind}. \texttt{Subfind} first uses the friend-of-friends (FOF) algorithm to find connected structures. We employ a linking length, $b=0.2$. Within FOF groups, \texttt{Subfind} maps the local density at all particle positions, finding the local maxima of the density distribution. Bound subhalos within FOF groups must consist of a least 30 particles.

Among the substructure detected within the virial radius of our simulated dwarf galaxies we identify four categories. Firstly, bound systems comprised solely of dark matter, which we can refer to as \textit{dark halos}. Secondly, objects that are and always have been dark matter dominated throughout their evolution but formed and continue to host stars. These are \textit{satellites}. Third, systems that formed stars within dark matter halos in the early Universe but that are now mostly comprised of stars. Their dark matter fraction at the present day is $f_\mathrm{DM}<50\%$, and in some cases zero. We will discuss their evolution in more detail in the following, suffice it to say here that their dark matter is stripped upon infall into larger dwarf galaxies. Hence, we will designate them \textit{stripped halos}. Finally, there are objects that formed directly as bound stellar systems, without ever having significant dark matter associated with them. These we will identify as \textit{star clusters}.

Of the total number of bound substructures that \texttt{Subfind} detects within the virial radius at $z=0$ between 95-99\% are dark halos. In turn,  $\sim99\%$ of the luminous structure is satellite galaxies. Finally, the luminous structure that is stellar dominated (i.e. with DM fractions less than 50\%) is of order 10-20. We note that not all the luminous structures we count here are expected to survive to $z=0$. See Sec.~\ref{sec:evaporation} for details. Finally, in the following we define ``infall'' as the first time the object moves within the virial radius of the host halo.

\subsection{Stripped halos at z=0}

  \begin{figure*}
    \centering
    \includegraphics[width=0.6\textwidth]{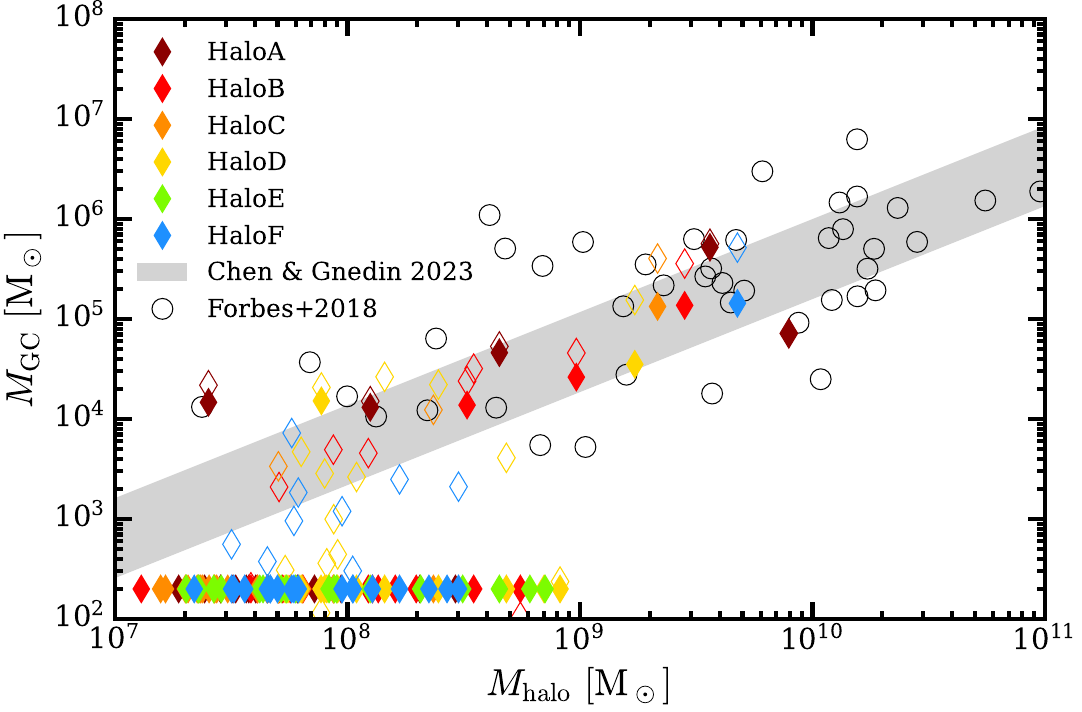}
    \caption{Total stellar mass in globular clusters as function of the halo (dynamical) mass of various main galaxies within the zoom region in each simulation. Empty black circles are data from \cite{Forbes2018}. The grey band is the fit to observational data by \cite{Chen2023}. The slope of the simulated points match the trend seen in observations very well.}
    \label{fig:gcmass}
  \end{figure*}

This paper centers on the investigation of stripped halos and their relationship with present-day globular clusters. In Fig.~\ref{fig:gridmass}, we present four illustrative cases demonstrating the evolution of mass over time for stripped halos. The dark matter (DM) mass is depicted in black, while the stellar mass is shown in red. The grey horizontal dashed lines denote the estimated timing of pericenter passages of the cluster into the main halo. It is evident that these clusters exhibit a dominance of dark matter at redshifts $z \gtrsim2$, typically possessing halo masses of $\Mhalo\gtrsim10^6\Msun$. The leftmost case experiences stripping during infall into the main halo, evident from the coinciding pericenter passages and reduction in DM mass. In contrast, the other three encounter stripping prior to their integration into the main halo. By $z=0$, the DM mass fractions fall well below 50\%, as indicated in the legend. Moreover, the stellar mass remains nearly constant throughout the evolution, with only minor mass loss, resulting in reductions of the total stellar mass by factors of 2-3 at most. We will delve into a detailed analysis of this behavior in the subsequent section.

Fig.~\ref{fig:gridmass} further underscores that while stripping predominantly affects the DM component, the stellar mass remains remarkably stable. The DM extends out to the virial radius, whereas the stars are concentrated in the central few parsecs. Tidal interactions, such as mergers, initially target the outer, less tightly bound particles. Eventually, in the central regions, the DM experiences preferential stripping due to its higher dynamical temperature compared to the star particles. This phenomenon is highlighted in Fig.\ref{fig:sigmaz5}. We calculate the velocity dispersion for all particles (separately for DM and stars) within the virial radius using the equation:
\begin{equation}
    \sigma = \sqrt{\frac{1}{3}\left(\sigma_x^2+\sigma_y^2+\sigma_z^2\right)}
\end{equation}
Across all cases, dark matter exhibits a greater velocity dispersion compared to the stars. The lower velocity dispersion in the stars arises from the gas kinematic properties of the natal cloud. Since gas can undergo cooling, condensation, and angular momentum loss, stars formed from this material are dynamically colder than the DM. After billions of years, what remains are stellar-dominated clusters with very low velocity dispersions ranging from $1-5\,\mathrm{km s}^{-1}$. Nevertheless, as demonstrated in \cite{Gutcke2022b}, the stellar velocity dispersion of the primary dwarf galaxies aligns with observed values for Local Group dwarfs. These central dwarfs build their mass from many minor and a few major mergers. Thus, the increase in velocity dispersion of the main galaxies is attributed to the accretion and merging of numerous small mini-halos, where their individual velocities contribute to the total dispersion, ultimately yielding properties consistent with dSph galaxies.

\subsection{Estimating the number of surviving GCs}
\label{sec:evaporation}

The DM and stars in our simulations are modeled as a collisionless fluid, approximating the density distribution of a single particle as a Dirac $\delta$-function convolved with a gravitational softening kernel. This approach softens or damps gravitational forces on scales below the softening length, $\varepsilon$. While the exact value of $\varepsilon$ is not crucial, it is imperative to ensure a sufficiently smooth and anisotropy-free field generated by individual particle positions. This N-body method, utilized in most cosmological codes, has proven computationally efficient and accurate at radii greater than the softening kernel.

However, as we advance towards higher resolutions and model individual stars, this collisionless treatment introduces inaccuracies affecting substructure counts. To address these omitted interactions, models like the one presented in \cite{McLaughlin2008} seek to include the evaporation in subgrid models.

Here, we will briefly discuss various processes that can lead to the dissolution of clusters and how well they might be captured in our simulations. Due to the high density of stars in cluster centers, stars are expected to undergo close encounters and collisions, known as "two-body interactions." Yet, collisionless star particles do not accurately represent the density distribution of the cluster on scales below the softening length. This disparity is most pronounced at the cluster centers, where artificially reduced densities emerge, akin to the artificial DM cores formed in collisionless DM simulations \citep[e.g.,][]{Tollet2016, DiCintio2014}. The gravitational forces pulling stars towards the center are damped, preventing radii from falling much below the softening scale. Consequently, structures are artificially inflated to a size no less than the softening length.

Furthermore, two-body interactions lead to an additional phenomenon -- two-body relaxation \citep{Spitzer1987}. This process involves clusters losing stars due to the heightened number of close encounters, resulting in the diffusion of stars across the tidal boundary. Due to the collisionless treatment of stars in our simulations, this evaporation is not captured. Consequently, once formed, star clusters persist even if they should have evaporated. Therefore, the number of clusters per halo in our simulation must be regarded as an upper limit.

The dissolution timescale for star clusters was estimated using N-body calculations of multi-mass star clusters subjected to an external tidal field \citep[equation 7]{Baumgardt2003}:
\begin{equation}
T_\mathrm{Diss} [\mathrm{Myr}] = \beta \left(\frac{N}{\ln(\gamma N) }\right) ^x \left(\frac{R_G}{\mathrm{kpc}} \right) \left( \frac{V_G}{220 \mathrm{km~s}^{-1}}\right)^{-1}
\end{equation}
where $N$ is the number of stars in the cluster, $\gamma$ is the Coulomb logarithm, $V_G$ is the circular velocity of the galaxy, and $R_G$ is the distance of the cluster from the galactic center. Here, $\gamma=0.02$ is a correction constant derived from models simulating clusters with a specific mass spectrum \citep{Giersz1996}, and $x$ is an exponent, that sets the relative importance of relaxation and tidal forces for the dissolution.

This equation assumes that each cluster resides within the tidal field of a galaxy from its inception (as opposed to falling in after formation) and that each cluster is a pure star cluster devoid of any associated dark matter. Both of these conditions do not align with our simulations entirely. Nonetheless, most clusters fall into the halo before $z=2$, meaning they spend over 10 billion years within the main potential. The third assumption -- that the larger potential remains constant -- is unrealistic, as the main halo of the dwarf experiences growth through accretion and undergoes mergers.

Following mass loss due to stellar evolution, the mass loss rate is anticipated to be linear in time for pure stellar systems. However, our clusters are not pure stellar systems until their dark matter is stripped. Thus, the mass loss rate is overestimated at early times. The significantly larger dark matter potential shields the clusters from tidal disruption, leading to preferential stripping of dark matter. Only after this phase is the stellar mass also stripped, as observed in Fig.~\ref{fig:gridmass}. 
While it would useful to interpret this deviation from linearity as an indication of the mass loss neglected in our simulations due to the use of collisionless dynamics, the mismatched assumptions do not allow this. Instead, we must assume that stellar mass loss and tidal disruption are sufficiently captured in the simulations and that two-body relaxation is the main missing process.

In the subsequent analysis, we thus estimate the missing two-body evaporation using the relaxation time definition \citep{Spitzer1987, Gnedin1997}:
\begin{equation}
\label{eq:trelax}
    t_{rh} = 0.138 \frac{M^{1/2}R^{3/2}}{G^{1/2}m_\star \ln(\gamma N)},
\end{equation}
where $M$ is the total cluster mass, $R$ is the half-mass radius, $m_\star$ is the average stellar mass, $\gamma$ is a correction constant as above, and $N$ is the number of stars in the cluster. We assume a cluster would dissolve due to two-body evaporation if 20 relaxation times have transpired since the cluster was predominantly composed of stars (see Fig.~\ref{fig:gridmass}). This time is generally between 1 and 6 Gyr.

\begin{figure*}
    \centering
    \includegraphics[width=0.7\textwidth]{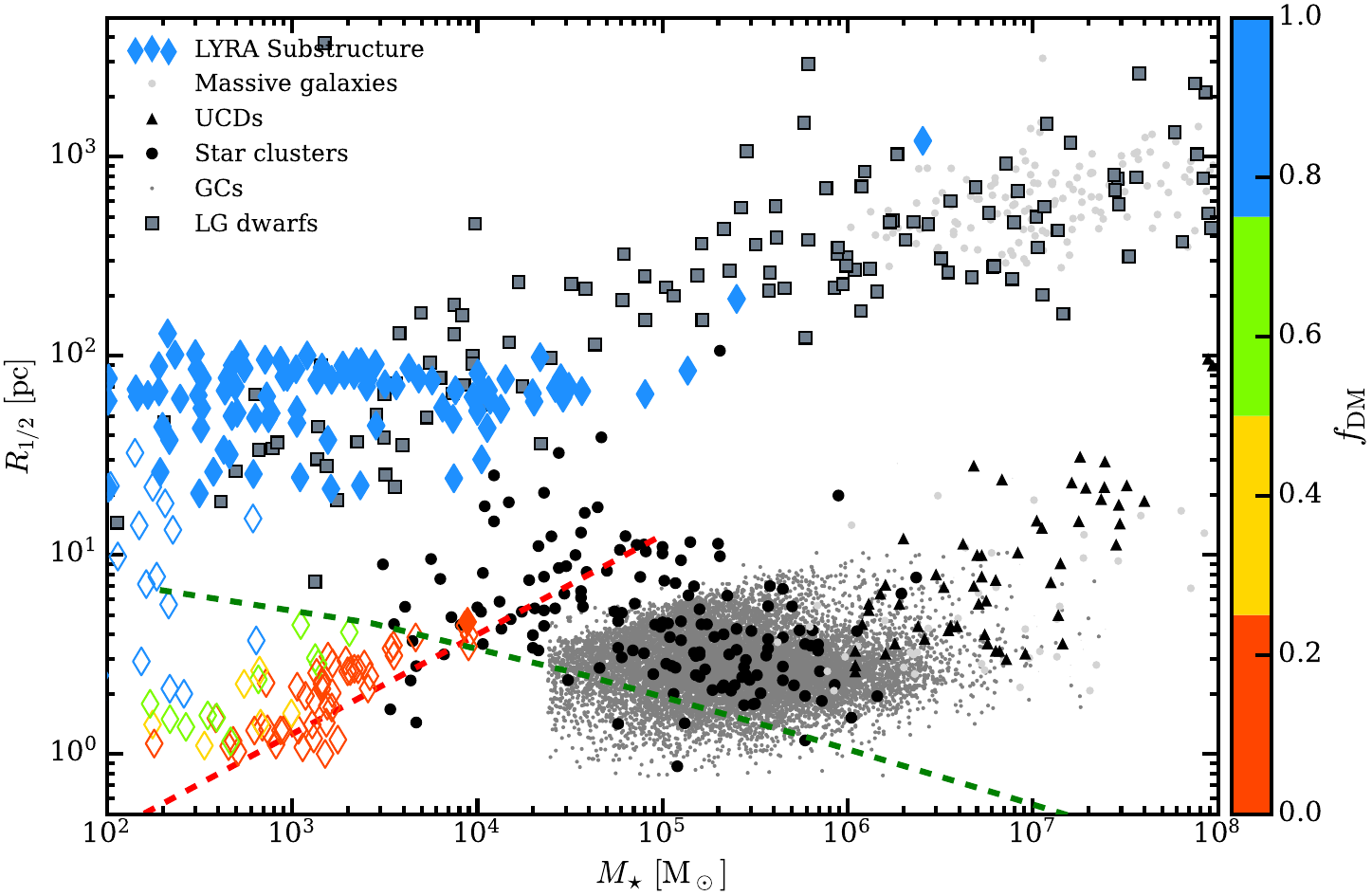}
    \caption{Size - mass relation of the high-resolution run (HaloF). Diamonds are substructure, the colors show the DM fraction within each subhalo, according to the colorbar. Black and grey symbols show Local Group dwarf galaxies, star clusters and globular clusters, according to the legend. 
    The red dashed line shows a line of constant stellar surface density, $\Sigma_\star\propto M^{0.5}=100~\Msun~\mathrm{pc}^{-2}$. Our simulated substructure covers the range between ultra-faint dwarf galaxies and star clusters. Of particular note is the single filled red diamond -- the single surviving GC at the present day.}
    \label{fig:size-mass}
  \end{figure*}

\subsection{GC mass -- halo mass relation}

In Fig.\ref{fig:gcmass}, we present the GC mass - halo mass relation. To ensure a broad comparison across various systems, we not only showcase the primary halo in each simulation but also include the ten most massive FOF groups within the high-resolution region. For those systems lacking GCs, we denote them with a value of 200. Among the remaining, we exclusively consider subhalos with dark matter fractions below 50\%. Additionally, we estimate the GCs that might have undergone evaporation as outlined in the preceding section. The summation of the remaining GCs, which are predominantly single objects, is denoted by filled diamonds in Fig.\ref{fig:gcmass}. The colors of these markers correspond to their respective simulations.

For clarity, the empty diamonds represent the total GC mass without subtracting the estimated evaporation. The open black circles denote individual dwarf galaxies measured by \cite{Forbes2018}, while the grey band signifies the fit to the GC mass - halo mass relation from \cite{Chen2023}. On a fundamental level, our simulations exhibit agreement with both the slope and normalization observed in reality. Upon comparing the filled and open diamonds, it becomes evident that the smallest GCs make only a marginal contribution to the overall GC mass. However, at the lowest halo masses, the relation experiences truncation, indicative of evaporation removing these GCs and consequently setting the total value on the $y$-axis to zero.

 \subsection{Size--mass relation}
 \label{sec:size}
Next, we delve into the properties of the stripped halos, beginning with an examination of the size-mass relationship. This correlation offers valuable insights into the density and classification of celestial objects, a topic thoroughly explored by \cite{Misgeld2011}. On the size-mass plane, both massive star clusters and ultra-faint dwarf galaxies share a common space. This has prompted the notion that certain star clusters may be the nuclei of stripped dwarf galaxies. However, present-day dwarf galaxies possess stellar velocity dispersions that are too high to give rise to nuclei of sufficient density to align with globular clusters. Instead, proto-galaxies or mini-halos at high redshifts can produce dense, low-velocity dispersion stellar systems, which can either merge to form dwarf galaxies or undergo stripping and persist as globular clusters.

The mass function of GCs is centered around ${2\times10^5\,\Msun}$, exhibiting a turnover at approximately ${4.4\times10^4\,\Msun}$. The surviving GCs in our galaxies predominantly occupy this lower mass range, with masses of around $10^4~\Msun$. Intriguingly, some known systems host GCs with very low masses. Limited observations of low-mass galaxies suggest a potential preference for hosting these lower-mass GCs, in line with the GC mass -- halo mass relation. Examples such as Eridanus II, Andromeda I, and Andromeda XXV exemplify this trend. Eridanus II, originally discovered by the Dark Energy Survey, has an estimated stellar mass of approximately $5.9\times10^4~\Msun$, with its GC near the center weighing in at $4\times10^3~\Msun$ \citep{Crnojevic2016}.

In Fig.~\ref{fig:size-mass}, we present the size--mass relation, where filled diamonds represent the remaining substructures post-evaporation calculation. Open diamonds encompass all identified substructures. The colormap on the right indicates the DM fraction, such that red symbols denote stellar--dominated systems. Black and grey symbols denote observational measurements of various celestial objects, as outlined in the legend. Specifically, we show the observational measurements from \cite{Misgeld2011}, which include star clusters \citep{McLaughin2005}, globular clusters \citep{Jordan2009}, and ultra-compact dwarf galaxies \citep[UCDs,][]{Mieske2008}. The grey squares are Local Group dwarfs from \cite{McConnachie2012}, where we estimated the masses using the fixed mass-to-light ratio of $M/L_V=1.8$.

We also plot two lines from analytic models. The green dashed line signifies the threshold where 20 relaxation times equate to 12 Gyr (cf. Eq.~\ref{eq:trelax}). Objects below this line are anticipated to dissolve within 12 Gyr. The red dashed line represents a constant surface density line, $\Sigma_\star \propto M^{0.5} = 100\Msun~\mathrm{pc}^{-2}$. As we see, this galaxy hosts a single remaining GCs at the present day (filled red diamond), which is also the most massive of the stripped halos.

\begin{figure*}
    \centering
    \includegraphics[width=\textwidth]{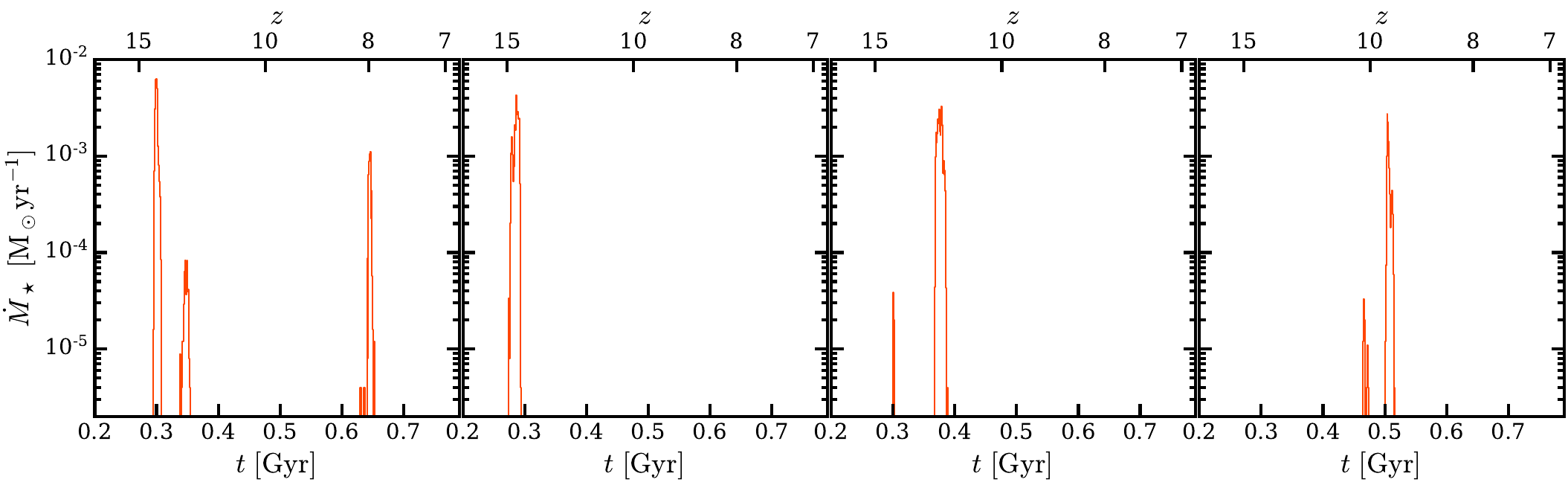}
    \includegraphics[width=\textwidth]{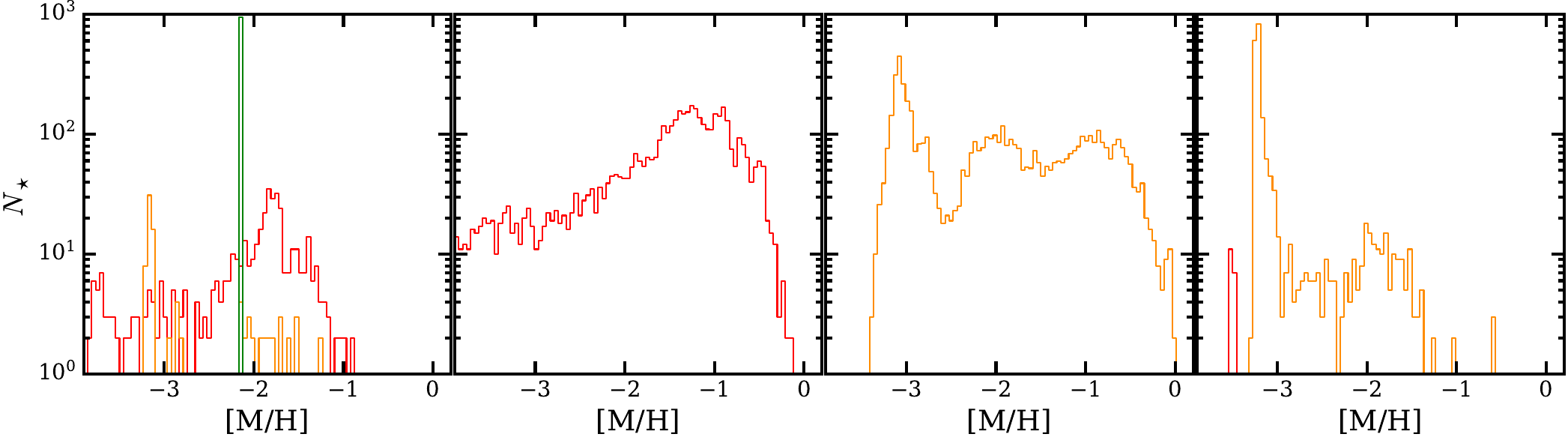}
    \caption{\textit{Top:} The star formation histories of four stripped halos. Some stripped halos show a single burst of star formation sometime between $20>z>8$. However, some stripped halos display multiple bursts of star formation, indicating multiple distinct stellar populations. \textit{Bottom:} Metallicity distribution functions for the stars within each stripped halo at $z=0$. Colors indicate the different burst of SF in time, from red to orange and then green. The width of the distribution is generally created within the single main burst rather than building up with consecutive bursts.}
    \label{fig:sfh}
  \end{figure*}

It is worth noting that our simulated galaxies also host a significant number of satellite galaxies, depicted as blue diamonds in Fig.~\ref{fig:size-mass}. While these objects do not serve as the primary focus of our present study, we would like to highlight a few noteworthy observations. Their placement on the size-mass relation situates them within the range of ultra-faint dwarf galaxies (UFDs). This is of interest, as there are indications that some of the UFDs identified within the Milky Way may have originated as companions to larger dwarf satellites like the Large and Small Magellanic Clouds \citep[e.g.][]{Cerny2023}. Additionally, it appears evident that there might be an excess of these satellites in our simulation. This could be attributed to the Population III star model implemented in our simulations \citep[see][]{Gutcke2022a}. Alternatively, it is possible that additional destruction mechanisms are not sufficiently captured in our model. A comprehensive analysis of these systems will be the subject of a forthcoming study and lies beyond the scope of our current work.

  \begin{figure}
    \centering
    \includegraphics[width=\columnwidth]{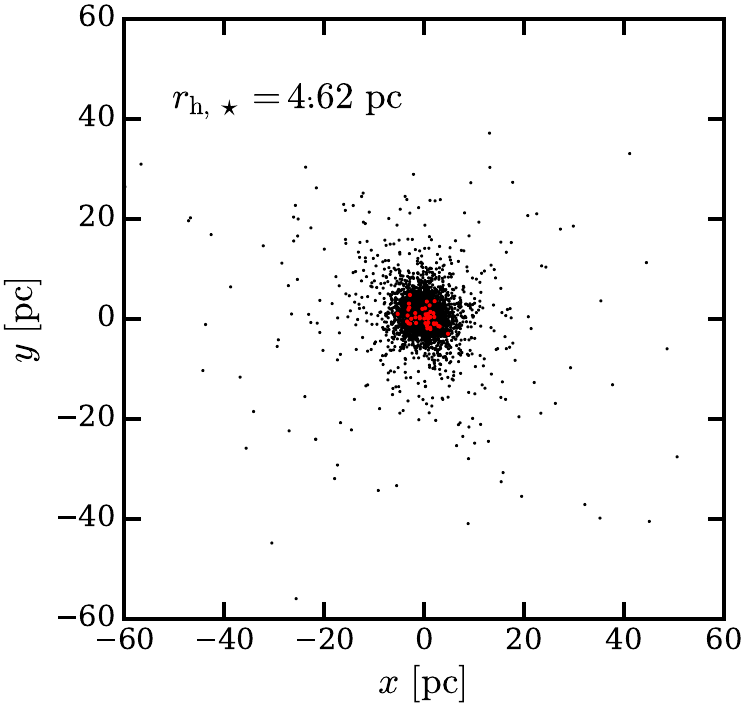}
    \caption{Distribution of stars in the single surviving GC from our high resolution run. The black dots are low mass surviving stars, while the red dots show massive remnants -- in this case stellar--mass black holes. While the remnants seem to be more centrally concentrated, this is simply an effect of low number statistics. We caution that mass segregation is not well captured on these scales by our collisionless dynamics.}
    \label{fig:image}
\end{figure}

  \subsection{Star formation histories and metallicity}

We turn to the individual star formation histories (SFHs) of the stripped halos, spotlighting the same selected four systems from Fig.~\ref{fig:gridmass}. The upper panel of Fig.~\ref{fig:sfh} presents the initial 800 Myr of cosmological time, roughly spanning $z\approx20-7$. Among these four systems, three exhibit distinct bursts of star formation, evoking the presence of multiple stellar populations akin to those found within globular clusters. Notably, all stars form prior to Reionization, attaining peak star formation rates of a few times $10^{-3}\mperyr$. 

The LYRA model tracks individual stars with masses exceeding $4~\Msun$, monitoring their evolution. These massive stars ultimately undergo various fates, transforming into asymptotic giant branch (AGB) stars, experiencing supernova events, or collapsing directly into black holes. Upon reaching the end of their lifecycles, the winds and ejecta from these stellar events are expelled into the surrounding interstellar medium (ISM). However, a residual mass persists, manifesting as a stellar remnant within the simulation. Consequently, we can identify these remnants within the GCs. 

In our high-resolution run, as presented in Fig.~\ref{fig:size-mass}, the single GC harbors 42 massive stellar remnants. At $z=0$, their masses fall within the range of $9-16\Msun$, whereas their initial masses spanned from $15-96~\Msun$. The spatial distribution of all stars within this GC is illustrated in black, with the massive remnants highlighted in red in Fig.~\ref{fig:image}. It is important to note that mass segregation, a process whereby heavier stars tend to concentrate toward the center of the cluster, is not effectively captured by collisionless dynamics, particularly within the softening length.

Turning our attention to the lower panel of Fig.~\ref{fig:sfh}, we encounter the corresponding metallicity distribution functions (MDFs). The various colored lines depict the metallicity distribution of stars in the different bursts (in burst order: red, orange, green). In general, the MDFs do not readily segregate based on stellar age. Instead, a broad MDF can emerge, even within a single burst, yielding stars with metallicities approaching solar values.

\section{Discussion \& Caveats} \label{sec:discussion}

It is interesting to use these simulations to predict what these dwarfs look like around $z\approx7-6$ where JWST may detect them and their substructure. Presently, even the central galaxies in our simulations at these redshifts tend to possess halo masses of $M_\mathrm{halo}\leq10^8~\Msun$ and stellar masses of $M_\star\leq10^6~\Msun$. Detection thresholds currently stand around 1.5 to 2 orders of magnitude higher, roughly at $M_\star\approx10^8~\Msun$, although lensing magnification can reduce this \citep[e.g.][]{Mowla2022}.

While making precise predictions for detecting GC-like substructures via JWST remains challenging, we can assert in general terms that if GC formation indeed necessitates stripped dark matter halos, dwarf galaxies detected with JWST in higher redshift ranges should exhibit significantly more substructure associated with them. This is consistent with findings for more massive galaxies, as demonstrated by \cite{Suess2023}, who detected numerous low-mass companions to quiescent galaxies out to $z<3$. Indeed, if GCs tend to evaporate and dissolve over time as modelled by \cite{Baumgardt2003}, we would expect to observe a far greater number of GCs in the early Universe compared to the count in the local Universe.

While the presented results depend on specific modeling choices, we anticipate that the fundamental physical processes underlying the formation of these structures will remain robust. However, there are several caveats to consider. Quantitative aspects like the number of bound objects in the present day or their exact masses are somewhat less secure and are contingent on specific model parameters. Furthermore, forthcoming work will extend the simulations to encompass more massive galaxies, potentially revealing more numerous and more massive GC-like objects.

The current simulation model employs the solar metallicity yield table outlined in \cite{Sukhbold2016}. A comprehensive examination of the gas phase properties in our simulations and the implications of solar metallicity yields will be detailed in Donaghue et al. (in preparation). As noted by \cite{Chieffi2004}, testing the metallicity dependency of supernova yields indicates that a straightforward scaling of solar metallicity abundances suffices to yield results within a factor of two. Indeed, the exact abundance ratios bear minimal influence on the simulations at runtime, as cooling rates are based solely on total metallicity, not individual elements.

More importantly, although supernovae have been identified as the primary driver of feedback and energy input in dwarf galaxies \cite[e.g.][]{Hu2017}, radiation from young stars also exerts a significant impact and is currently not integrated into our simulations. Radiation acts earlier than the first supernova events, creating HII regions and rarefying the gas in the natal cloud. This amplifies the feedback effect of subsequent supernovae and may truncate star formation in early mini-halos, resulting in lower stellar mass systems than those presented in this study. As our model evolves, we plan to incorporate this process, providing a definitive assessment of its impact on our conclusions.

\section{Conclusions} \label{sec:conclusion}

In this study, we present a formation scenario for present-day globular clusters (GCs) as stellar remnants of stripped dark matter halos. Through a series of zoom-in cosmological simulations using the LYRA model, we explore the evolution of these globular cluster--like objects across cosmic time. Our findings offer insights into the processes underlying the birth and survival of GCs.

One of the key results of our work lies in their mass evolution. The structures transition from dark matter-dominated objects at high redshifts ($z \gtrsim2$) to being comprised of substantial stellar mass at $z=0$. This transformation occurs due to tidal stripping during infall into larger halos and underscores the dynamic past of these systems. Due to this history, the simulations naturally reproduce the total GC mass--halo mass relation, even at these low masses.

We also present the star formation histories and metallicity distribution functions, showing multiple bursts reminiscent of the multiple stellar populations commonly observed in GCs. These are more easily understood in our stripped halo scenario, since the objects are hosted within DM halos at the time of their star formation and can re-accrete gas, giving rise to distinct star formation episodes. We additionally show that massive stellar remnants are hosted in the centers of the simulated GCs.

The convergence of massive star clusters and ultra-faint dwarf galaxies in the size-mass parameter space has prompted consideration of stripped dwarf galaxies as potential progenitors of GCs. Our simulations connect these two populations directly. This additionally finds support in the dense, low-velocity dispersion stellar systems characteristic of high-redshift proto-galaxies.

Our work offers implications for observations with JWST. While direct predictions regarding the detection of GC-like substructures remain challenging, our simulations suggest that JWST observations of dwarf galaxies at higher redshifts should reveal a markedly heightened prevalence of associated substructure. This aligns with recent findings pertaining to more massive galaxies.
As our understanding of GC evolution continues to develop, we anticipate that future studies will further refine and expand upon the insights presented here.
Future extensions of this work to encompass more massive galaxies promise to yield further insights into the formation of GC-like structures.

\section*{Acknowledgments}
TAG acknowledges support by NASA through the NASA Hubble Fellowship grant $\#$HF2-51480 awarded by the Space Telescope Science Institute, which is operated by the Association of Universities for Research in Astronomy, Inc., for NASA, under contract NAS5-26555.
We acknowledge the computing time provided by the Leibniz Rechenzentrum (LRZ) of the Bayrische Akademie der Wissenschaften on the machine SuperMUC-NG (pn73we).
This research was also carried out on the High Performance Computing resources of the FREYA and COBRA clusters at the Max Planck Computing and Data Facility (MPCDF, \url{https://www.mpcdf.mpg.de}) in Garching operated by the Max Planck Society (MPG).

\bibliography{bib}{}
\bibliographystyle{aasjournal}


\end{document}